\documentclass[showpacs,preprintnumbers,amsmath,aps,nofootinbib,amssymb,superscriptaddress]{revtex4}
\usepackage{graphicx}% Include figure files
\usepackage[english]{babel}
\usepackage{bm}% bold math
\usepackage{enumerate}
\usepackage[tikz]{bclogo}
\usepackage[framemethod=tikz]{mdframed}

\newsavebox{\measurebox}

\begin{document}
\baselineskip 16pt
\title{Classical and quantum exact solutions for a FRW multi-scalar field cosmology with an exponential potential driven inflation}
\author{J. Socorro}
\email{socorro@fisica.ugto.mx}
\author{Omar E. N\'u\~nez}
\email{neophy@fisica.ugto.mx}
\affiliation{Departamento de  F\'{\i}sica, DCeI, Universidad de Guanajuato-Campus Le\'on, C.P. 37150, Le\'on, Guanajuato, M\'exico}
\author{Rafael Hern\'{a}ndez-Jim\'{e}nez}
\email{s1367850@sms.ed.ac.uk}
\affiliation{School of Physics and Astronomy, University of Edinburgh, Edinburgh, EH9 3FD, United Kingdom}

\begin{abstract}

A flat Fiedmann-Robertson-Walker (FRW) multi-scalar field cosmology is
studied with a particular potential of the form $ \rm V(\phi,\sigma)=V_0
e^{-\lambda_1 \phi-\lambda_2 \sigma}$, which emerges as a relation between the time derivatives of the scalars field momenta. Classically, by employing the Hamiltonian formalism
of two scalar fields $\rm(\phi,\sigma)$ with standard kinetic energy,
exact solutions are found for the Einstein-Klein-Gordon (EKG)
system for different scenarios specified by the parameter $\rm\lambda^2=\lambda_1^2+\lambda_2^2$, as well as the e-folding function $\rm N_{e}$ which is also computed.
For the quantum scheme of this model, the corresponding Wheeler-DeWitt (WDW) equation is solved by applying an appropriate change of variables.
\end{abstract}
\pacs{4.20.Fy, 4.20.Jb, 98.80.-k, 98.80.Hw}
\maketitle

\section{Introduction}

The inflation paradigm is considered the most accepted mechanism to explain many of the fundamental problems of the early stages in the evolution of our universe \cite{guth1981, linde1982, turner1981, starobinsky1980}, such as the flatness, homogeneity and isotropy observed in the present universe. Another important aspect of inflation is its ability to correlate cosmological scales that would otherwise be disconnected. Fluctuations generated during this early phase of inflation yield a primordial spectrum of density perturbation \cite{Starobinsky:1979ty,Mukhanov:1981xt,kodama, bassett}, which is nearly scale invariant, adiabatic and Gaussian, which is in agreement with cosmological observations \cite{Planck}.

The single-field scalar models have been broadly used to describe the primordial expansion, the most phenomenological successful are those with a quintessence scalar field and slow-roll inflation \cite{barrow, andrew1998b, ferreira, copeland1, copeland2, copeland3, andrew2007, gomez, capone, kolb}. However, if another component is included, i.e. a multi-scalar field theory, it is also possible to produce an inflationary scenario  \cite{coley, Copeland:1999cs}, even if the fields are non interacting \cite{andrew2007}. Even more the dynamical possibilities in  multi-field inflationary scenarios are considerably richer than in single-field models, such as in the primordial inflation perturbations analysis \cite{Yokoyama:2007dw, Chiba:2008rp} or the assisted inflation as discussed in \cite{andrew1998a}, furthermore, the general assisted inflation as in \cite{Copeland:1999cs}. In this sense the multi-scalar fields cosmology is an attractive candidate to explain such phenomenon.

Recent works have shown that multi-scalar field models are very fruitful when studying the early stages of the universe, such is the case in \cite{DeCross}, where the authors perform a semi-analytic study of preheating in inflationary models comprised of multiple scalar fields coupled nonminimally to gravity. In \cite{Hotinli:2017vhx} the authors the sensitivity of the cosmological observables to the reheating phase following inflation driven by many scalar fields, where they find that for certain decay rate, reheating following multi-field inflation can have a significant impact on the prediction of cosmological observables.

Indeed the multi-scalar field models for inflation are of interest even on most recent studies, such as the above mentioned cases, however, one of the most important features in such models is the potential associated to the scalar fields, and in many cases, the employed potentials are simple polynomial powers of the scalar fields or in other cases the employed potential is a series of lineally summed  exponentials, however, it has been shown that a potential of the form $\rm V(\phi,\sigma)=V_0 e^{-\lambda_1 \phi -\lambda_2 \sigma}$  is a good candidate to model the inflation phenomenon for multi-scalar field theory, as discussed in previous work \cite{omar-epjp2017}, and might provide a richer post inflation scenario.

Generally, in the studies of inflationary cosmology one employs the usual slow-roll approximation with the objective to extract simple expression for basics observable, such as the scalar and tensor spectral indices, the
running of the scalar spectral index and the tensor-to-scalar ratio.
Moreover, in the slow-roll regime the set of EKG equations reduces in such a way that one can quickly obtain the solution of the scale factor.
Nevertheless, there is an alternative approach which allows for an easy derivation of many inflation results. It is called the Hamilton's formulation, widely used in analytical mechanics.
Using this method we obtain the exact solutions of the complete set of EKG equations without using the aforementioned approximation.

On the other hand, we implement a basic formulation in quantum cosmology by means of the Wheeler-DeWitt (WDW) equation. The WDW equation has been analyzed with different approaches in order to solve it, and there are several papers on the subject, such is the case in \cite{Gibbons}, where they debate what a typical wave function for the universe is. In ref. \cite{Zhi} has a review on quantum cosmology where the problem of how the universe emerged from big bang singularity can no longer be neglected in the GUT epoch. Moreover, the best candidates for quantum solutions are those that have a damping behavior with respect to the scale factor, since only such wave functions allow for good classical solutions when using a Wentzel-Kramers-Brillouin (WKB) approximation for any scenario in the evolution of our universe \cite{HH,H}. Furthermore, in the context of a single scalar field a family of scalar potentials is obtained in the Bohmian formalism \cite{wssa,nuevo}, where among others a general potential of the form $\rm V(\phi)=V_0 e^{-\lambda \phi}$ is examined. Given this insight, for a two scalar field scenario we consider a potential of the form $\rm V(\phi,\sigma)=V_0 e^{-\lambda_1 \phi -\lambda_2 \sigma}$ in order to solve the WDW equation.

\bigskip

This work is arranged as follows. In section \ref{model} we present the model with the action and the corresponding EKG equations for our cosmological model and the associated Hamiltonian density. In section \ref{csolutions}  general equations for the classical solutions of scale factor, scalar fields and their associated momenta are derived in terms of the free parameters of the model. In subsections  \ref{csolutionsa}, \ref{csolutionsb}, \ref{csolutionsc} and \ref{csolutionsd} the particular solutions and their number of e-folds is computed  for different cases of the $\rm \lambda$ parameter. in section \ref{qsolutions} we use the Hamiltonian density to compute the corresponding WDW equation, which is solved by using a change of variables, an ansatz for the wave function is employed in terms of a generic function and parameters which are to be determined. In subsections \ref{qsolutionsa} and \ref{qsolutionsb} the corresponding wave function and their constants relations are presented for different cases of the $\rm \delta$ parameter, which in turn is related to the $\rm \lambda$ parameter of the classical solutions. Finally, in section \ref{conclusions} we present our conclusions for this work.

%%%%%%%%%%%%%%%%%%%%%%%%%%%%%%%%%%%%%
%%%%%%%%%THE MODEL%%%%%%%%%%%%%%%%%%%
%%%%%%%%%%%%%%%%%%%%%%%%%%%%%%%%%%%%%

\section{The model \label{model}}

We begin with the construction of two scalar fields cosmological paradigm, which requires canonical scalar fields $\rm \phi,\sigma$. The action of a universe with the constitution of such fields is
\begin{equation}
\rm {\cal L}=\sqrt{-g} \left( R+\frac{1}{2}g^{\mu\nu}\nabla_\mu \phi \nabla_\nu \phi
+\frac{1}{2}g^{\mu\nu}\nabla_\mu \sigma \nabla_\nu \sigma -
V(\phi,\sigma)\right) \,, \label{lagra}
\end{equation}
where $\rm R$ is the Ricci scalar, $\rm V(\phi,\sigma)$ is the corresponding scalar field potential, and the reduced Planck mass $M_{P}^{2}=1/8\pi G=1$. The corresponding variations of Eq.(\ref{lagra}), with respect to the metric and the scalar fields give the Einstein-Klein-Gordon field equations
\begin{eqnarray}
\rm G_{\alpha \beta} = \frac{1}{2}\left(\nabla_\alpha \phi \nabla_\beta \phi -\frac{1}{2}g_{\alpha \beta} g^{\mu \nu}
\nabla_\mu \phi \nabla_\nu \phi \right) &+&
\frac{1}{2}\left(\nabla_\alpha \sigma \nabla_\beta \sigma
-\frac{1}{2}g_{\alpha \beta} g^{\mu \nu}
\nabla_\mu \sigma \nabla_\nu \sigma \right)-\frac{1}{2}g_{\alpha \beta} \, V(\phi,\sigma), \label{munu}\\
\rm \Box \phi -\frac{\partial V}{\partial \phi} = g^{\mu\nu} \phi_{,\mu\nu} &-& g^{\alpha \beta} \Gamma^\nu_{\alpha
\beta} \nabla_\nu \phi - \left(\frac{\partial V}{\partial \phi}\right)_\sigma=\rm
0 \,,\label{ekg-phi} \\
\rm \Box \sigma -\frac{\partial V}{\partial \sigma} = g^{\mu\nu} \sigma_{,\mu\nu} &-& g^{\alpha \beta} \Gamma^\nu_{\alpha
\beta} \nabla_\nu \sigma - \left(\frac{\partial V}{\partial \sigma} \right)_\phi = 0 \,. \label{ekg-sigma}
\end{eqnarray}
The line element to be considered in this work is the flat FRW
\begin{equation}
\rm ds^2=-N(t)^2 dt^2 +e^{2\Omega(t)} \left[dr^2
+r^2(d\theta^2+sin^2\theta d\phi^2) \right], \label{frw}
\end{equation}
where $\rm N$ is the lapse function, which in a special gauge one can directly recover the cosmic time $\rm t_{phys}$ ($\rm Ndt=dt_{phys}$), the scale factor $\rm A(t)=e^{\Omega(t)}$ is in the Misner's parametrization, and the scalar function has an interval, $\rm \Omega \in (-\infty,\infty)$.
Consequently the field equations are
\begin{eqnarray}
\rm \frac{3\dot{\Omega}^{2}}{N^2}-\frac{\dot{\phi}^2}{4N^2}-\frac{\dot{\sigma}^2}{4N^2}-\frac{1}{2}V(\phi,\sigma)&=&0 \,, \label{ein0}\\
\rm \frac{2\ddot{\Omega}}{N^2}+\frac{3\dot{\Omega}^2}{N^2}-\frac{2\dot{\Omega}\dot{N}}{N^3}+\frac{\dot{\phi}^2}{4N^2}+\frac{\dot{\sigma}^2}{4N^2}
-\frac{1}{2}V(\phi,\sigma)&=&0 \,, \label{ein1}\\
\rm \frac{\ddot{\phi}\dot{\phi}}{N^2}+\frac{3\dot{\Omega}\dot{\phi}^2}{N^2}-\frac{\dot{N}\dot{\phi}^2}{N^3}+\left(\dot V\right)_\sigma &=& 0 \,, \label{ein2}\\
\rm \frac{\ddot{\sigma}\dot{\sigma}}{N^2}+\frac{3\dot{\Omega}\dot{\sigma}^2}{N^2}-\frac{\dot{N}\dot{\sigma}^2}{N^3}+\left(\dot V\right)_\phi &=& 0 \,. \label{ein3}
\end{eqnarray}

By building the corresponding Lagrangian and Hamiltonian densities for this cosmological model, classical solutions to Einstein-Klein-Gordon Eqs.(\ref{munu}-\ref{ekg-sigma}) can be found using the Hamilton's approach, and the quantum formalism can be determined and solved. In that sense, we use the metric Eq.(\ref{frw}) into Eq.(\ref{lagra}) having
\begin{equation}\label{lagrafrw}
\rm {\cal{L}}= \rm e^{3\Omega}\left(\frac{6\dot{\Omega}^2}{N}-\frac{\dot{\phi}^2}{2N^2}-\frac{\dot{\sigma}^2}{2N^2}+N V(\phi,\sigma)\right)\,,
\end{equation}
where upper ``{\tiny{$\bullet$}}" represents the first time derivative, and the corresponding momenta are defined in the usual way $\rm \Pi_q=\partial {\cal L}/\partial \dot q$. We obtain
\begin{eqnarray}
\rm \Pi_\Omega &=& 12 \frac{e^{3\Omega}}{N}\dot \Omega \,,
\qquad\qquad \dot \Omega=\frac{N e^{-3\Omega}}{12} \Pi_\Omega \,, \nonumber\\
\rm \Pi_\phi&=& -\frac{e^{3\Omega}}{N}\dot \phi,\qquad\qquad \,\,
\dot \phi=-N e^{-3\Omega} \Pi_\phi \,, \label{momenta} \\
\rm \Pi_\sigma&=& -\frac{e^{3\Omega}}{N}\dot \sigma,\qquad\qquad \,\,
\dot \sigma=-N e^{-3\Omega} \Pi_\sigma \,. \nonumber
\end{eqnarray}
By performing the variation of the canonical Lagrangian with respect to $\rm N$, i.e. $\rm\delta{\cal L}_{canonical}/\delta N=0$, where $\rm{\cal L}_{canonical}= \Pi_q \dot{q}-N{\cal H}$, it implies the constraint $\rm{\cal H}=0$. Hence the Hamiltonian density is
\begin{equation}
\rm {\cal H}= \frac{e^{-3\Omega}}{24} \left[ \Pi_\Omega^2-12
\Pi_\phi^2-12\Pi_\sigma^2-24  V(\phi,\sigma) e^{6\Omega}\right] \,. \label{hamifrw}
\end{equation}
In the gauge $\rm N=24 e^{3\Omega}$ and using the Hamilton equations $\rm \dot{q}=\partial {\cal H}/\partial \Pi_{q}$ and $\rm \dot{\Pi}_q=-\partial {\cal H}/\partial q$, we have the following set of equations
\begin{eqnarray}
\rm \dot \Omega &=&  2 \Pi_\Omega, \qquad \dot \phi= -24 \Pi_\phi \,, \qquad \dot \sigma= -24 \Pi_\sigma, \nonumber\\
\rm \dot \Pi_\Omega&=&  6U \,, \,\,\,\,\quad \dot \Pi_\phi= \frac{\partial U}{\partial \phi} \,,
\,\,\,\,\,\qquad \dot \Pi_\sigma= \frac{\partial U}{\partial \sigma} \,,\label{new-variables}
\end{eqnarray}
where $\rm U=24V(\phi,\sigma)e^{6\Omega}$. Given a particular form of the potential $\rm V(\phi,\sigma)$ one can derive a relation between the time derivative of the momenta such as
 $\rm \dot{\Pi}_{\phi} \propto \dot{\Pi}_{\sigma}$, provided that $\partial V/\partial\phi=\alpha\partial V/\partial\sigma$, where $\alpha$ is a constant. Such connection can be obtained considering two different configurations of the potential: $\rm V(\phi,\sigma)=f[\pm(\alpha_1 \phi+\alpha_2 \sigma)]$ or $\rm V(\phi,\sigma)=V_1f[\pm(\alpha_1 \phi)] + V_2f[\pm(\alpha_1 \sigma)] $, where $\rm V_{1}$ and $V_{2}$ are constants, and $\rm f(\phi,\sigma)$ is an arbitrary function.
We select the simplest form of the potential $\rm V(\phi,\sigma)=A(\phi)B(\sigma)$:
\begin{equation}
\rm V=V_0 e^{-\lambda_1 \phi - \lambda_2 \sigma } \,,
\end{equation}
where $\rm V_{0}$ is a constants and $\rm \lambda_{1}$ and $\lambda_{2}$ are distinguishing parameters. This class of potential has been obtained by other methods, see for instance \cite{gssa,socorro-doleire,socorro-pimentel,ssw,omar-epjp2017}.
Therefore the time derivative of the momenta are simply $\rm \dot \Pi_\phi= -\lambda_1 U$ and
$\rm \dot \Pi_\sigma= -\lambda_2 U$, which solutions are
\begin{equation}
\rm \Pi_\phi=-\frac{\lambda_1}{6}\Pi_\Omega+p_\phi, \qquad
 \Pi_\sigma=-\frac{\lambda_2}{6}\Pi_\Omega+p_\sigma, \qquad   \label{momentos}
\end{equation}
where $\rm p_\phi$ and $\rm p_\sigma$ are integration constants. Henceforth we will employ this scheme in order to find analytic classic and quantum solutions.

%%%%%%%%%%%%%%%%%%%%%%%%%%%%%%%%%%%%%
%%%%%%%%%CLASSICAL SOLUTIONS%%%%%%%%%
%%%%%%%%%%%%%%%%%%%%%%%%%%%%%%%%%%%%%

\section{Classical solutions \label{csolutions}}

We start from the Hamilton equations Eq.(\ref{new-variables}) in order to find relations between the  scale factor and the scalar fields, such as
\begin{eqnarray}
\rm && \dot\phi = -24\Pi_\phi= 4\lambda_1 \Pi_\Omega -24p_\phi=2\lambda_1 \dot \Omega -24 p_{\phi} \,, \\
\rm && \dot\sigma = -24\Pi_\sigma= 4\lambda_2 \Pi_\Omega -24p_\sigma=2\lambda_2 \dot \Omega -24 p_{\sigma} \,,
\end{eqnarray}
which solutions are
\begin{eqnarray}
\rm && \phi=\phi_1+2\lambda_1 \Omega -24p_\phi t \,, \label{phi}\\
\rm && \sigma=\sigma_1+2\lambda_2 \Omega -24p_\sigma t \,, \label{sigma}
\end{eqnarray}
where $\rm\phi_{1}$ and $\rm\sigma_{1}$ are integration constants, and they be determined by suitable conditions. These expression are indeed general relations, since they satisfy the Einstein-Klein-Gordon equations Eqs.(\ref{ein0}-\ref{ein3}). Then by taking into account the constraint $\rm {\cal H}=0$, we obtain
the temporal dependence for $\rm \Pi_\Omega(t)$ which allows us to construct a master equation:
\begin{equation}
\rm \frac{d \Pi_\Omega}{m_1 \Pi_\Omega^2 + m_2 \Pi_\Omega - m_3}=dt \,,
\label{master-equation}
\end{equation}
where the parameters $\rm m_i\,,\,i=1,2,3$, are
\begin{equation}\label{parameter}
\rm m_1=2\left(3-\lambda_1^2-\lambda_2^2\right)=2(3- \lambda^2) \,,\quad m_2=24\left[\lambda_1 p_\phi+\lambda_2 p_\sigma \right] \,, \quad
m_3=72\left[p_\phi^2+ p_\sigma^2\right] \,.
\end{equation}
Subsequently by analyzing the parameter
$\rm\lambda^2=\lambda_1^2 + \lambda_2^2$ we will obtain three different solutions.

\subsection{Solution for $\rm\lambda^2=3$  \label{csolutionsa}}

Having $\rm\lambda^2=3$ implies that $\rm m_1=0$, so the integral to solve becomes
\begin{equation}
\rm \int\frac{d \Pi_\Omega}{m_2 \Pi_\Omega - m_3} = \int dt \,,
\end{equation}
then we parameterize $\lambda_i$ such as:
$\rm \lambda_1=\sqrt{3 (1-\epsilon)}\,,\,
\lambda_2=\sqrt{3\epsilon}\,,\, \lambda^2=3$, where $\epsilon
\in (0,1)$ measures the corresponding weight for each scalar field during inflation; so the constants Eq.(\ref{parameter}) become
\begin{equation}\label{n-parameter}
\rm m_2=24\sqrt{3(1-\epsilon)}\left[p_\phi+ \sqrt{\frac{\epsilon}{ (1-\epsilon)}} p_\sigma\right] \,,\quad m_3 = 72\left[ p_\phi^2+ p_\sigma^2\right] \,.
\end{equation}
Thus $\rm \Pi_\Omega(t)$ becomes
\begin{equation}\label{Pi-Omega-lambda2-equal-3}
\rm\Pi_\Omega(t)=\frac{m_3}{m_2} +c_1 e^{bt} \,,
\end{equation}
were $\rm c_1$ is an integration constant. Using the relations from Eq.(\ref{new-variables}) and after some algebra, the solutions of the set of variables $\rm(\Omega,\phi, \sigma)$ and
$\rm (\Pi_\phi,\Pi_\sigma)$ are:
\begin{eqnarray}
&&\rm \Omega=\Omega_0 +\frac{2m_3}{m_2}t+ \frac{2c_1}{m_2} e^{m_2t} \,,\\
&&\rm \phi=\phi_0 + 4\sqrt{3(1-\epsilon)}\frac{m_3}{m_2} t-24p_\phi t + \frac{4c_1\sqrt{3(1-\epsilon)}}{m_2} e^{m_2t} \,, \\
&&\rm \sigma=\sigma_0 + 4\sqrt{3\epsilon}\frac{m_3}{m_2} t -24 p_\sigma t + \frac{4 c_1 \sqrt{3\epsilon}}{m_2} e^{m_2t} \,, \\
&&\rm \Pi_\phi=-\frac{\sqrt{3(1-\epsilon)}}{6}\left(\frac{m_3}{m_2} + c_1 e^{m_2t}\right)+p_\phi \,, \\
&&\rm \Pi_\sigma=-\frac{\sqrt{3\epsilon}}{6} \left(\frac{m_3}{m_2} + c_1 e^{m_2t}\right) +p_\sigma \,,
\end{eqnarray}
where ($\rm\Omega_{0},\phi_{0},\sigma_{0}$) are all integration constants.
In order to above results fulfill the EKG Eqs.(\ref{ein0}-\ref{ein3}), all constants must satisfy that $\rm 144V_0=m_2 c_1e^{-6\Omega_0+\lambda_1 \phi_0+\lambda_2 \sigma_0}$.
Finally the scale factor $\rm A(t)$ for this case is
\begin{equation}\label{huge-scale}
\rm A=A_0 e^{\left(\frac{2m_3}{m_2}\right)t} \,Exp\left[ \frac{2c_1}{m_2} e^{m_2t}\right] \,,
\end{equation}
where $\rm A_{0}=e^{\Omega_{0}}$. Given that the scale factor is an exponential of an exponential function, it might exhibit a highly substantial growth.

\subsection{Solution for $\rm\lambda^{2}>3$ \label{csolutionsb}}

For this case $\rm m_1=2(3-\lambda^2)<0$, so the integral to solve becomes
\begin{equation}\label{master-integral-lambda-bigger3}
\rm \frac{d \Pi_\Omega}{-m_1 \Pi_\Omega^2+  m_2 \Pi_\Omega - m_3}=dt \,,
\end{equation}
where we include the minus sign in this equation, such the constant $\rm m_1=2(\lambda^2-3)=2\beta>0$. Then we define $\rm \omega^2 =m_2^2 -8\beta m_3$, so we change variable as $\rm z=4\beta \Pi_\Omega -m_2$ in order to integrate Eq.(\ref{master-integral-lambda-bigger3}). Thus the solution to the momenta $\rm \Pi_\Omega(t)$ becomes
\begin{equation}
\rm \Pi_\Omega=\frac{m_2}{4\beta}+\frac{\omega}{4\beta} tanh\left( \frac{\omega}{2}t \right) \,.
\end{equation}
Using the relations from Eq.(\ref{new-variables}) and after some algebra, the solutions of the set of variables $\rm(\Omega,\phi, \sigma)$ and
$\rm (\Pi_\phi,\Pi_\sigma)$ are:
\begin{eqnarray}
&&\rm \Omega=\Omega_0 + \frac{m_2}{2\beta}t + \frac{1}{\beta} Ln\left[cosh\left(\frac{\omega}{2}t\right)\right]  \,,\\
&&\rm \phi =\phi_0 + \left(\lambda_{1}\frac{m_2}{\beta}-24p_\phi\right) t -\frac{2\lambda_{1}}{\beta}\ln\left[cosh\left(\frac{\omega}{2}t\right) \right] \,, \\
&&\rm \sigma=\sigma_0 + \left(\lambda_{2}\frac{m_2}{\beta}-24p_\sigma\right) t -\frac{2\lambda_{2}}{\beta}\ln\left[cosh\left(\frac{\omega}{2}t\right) \right] \,, \\
&&\rm \Pi_\phi=-\frac{\lambda_{1}}{6}\left(\frac{m_2}{4\beta}+\frac{\omega}{4\beta} tanh\left( \frac{\omega}{2}t \right)\right)+p_\phi \,, \\
&&\rm \Pi_\sigma=-\frac{\lambda_{2}}{6} \left(\frac{m_2}{4\beta}+\frac{\omega}{4\beta} tanh\left( \frac{\omega}{2}t \right)\right) +p_\sigma \,,
\end{eqnarray}
where ($\rm\Omega_{0},\phi_{0},\sigma_{0}$) are all integration constants.
In order to above results fulfill the EKG Eqs.(\ref{ein0}-\ref{ein3}), all constants must satisfy that $\rm 1152 \beta V_0=\omega^{2}e^{\lambda_1 \phi_0 +\lambda_2 \sigma_0+2\beta\Omega_{0}}$.
Finally the scale factor becomes
\begin{equation}
\rm A=A_0\, e^{\frac{m_2}{2\beta}t}\, \left[cosh\left(\frac{\omega}{2}t\right)\right]^{\frac{1}{\beta}} \,,
\end{equation}
where $\rm A_{0}=e^{\Omega_{0}}$. For this case, given that $\rm\beta>0$, one would expect that the scale factor grows slower than the previous case $\rm\lambda^{2}=3$.

\subsection{Solution when $\lambda^2 < 3$ \label{csolutionsc}}

For this case we modify the relation between the momenta Eq.(\ref{momentos}), by changing the sign in the constants, ($\rm p_\phi,p_\sigma,m_2$)$\to$($\rm-p_\phi,-p_\sigma,-m_2$), and $\rm m_1=2(3-\lambda^2)=2\eta>0$; therefore the integral to solve becomes
\begin{equation}
\rm \frac{d\Pi_\Omega}{2\eta \Pi_\Omega^2 - m_2 \Pi_\Omega-m_3}=dt \,.
\end{equation}
Thus $\rm\Pi_\Omega(t)$ is
\begin{equation}
\rm \Pi_\Omega = \frac{1}{4\eta}\left[m_2-\alpha\coth\left(\frac{\alpha}{2}t\right)\right] \,,
\end{equation}
where $\rm \alpha^2=m_2^2+8\eta m_3$. Using the relations from Eq.(\ref{new-variables}) and after some algebra, the solutions of the set of variables $\rm(\Omega,\phi, \sigma)$ and
$\rm (\Pi_\phi,\Pi_\sigma)$ are:
\begin{eqnarray}
&&\rm \Omega = \Omega_0 +\frac{m_2}{2\eta}t+\ln{\left[csch{\left(\frac{\alpha}{2}t\right)}\right]}^{1/\eta} \,,\\
&&\rm \phi = \phi_0 + \left(\lambda_1 \frac{m_2}{\eta}+24p_\phi\right) t -\ln{\left[\sinh{\left(\frac{\alpha}{2}t\right)}\right]}^{2\lambda_1/\eta} \,, \\
&&\rm \sigma = \sigma_0 +\left( \lambda_2 \frac{m_2}{\eta}+24p_\sigma\right) t -\ln{\left[\sinh{\left(\frac{\alpha}{2}t\right)}\right]}^{2\lambda_2/\eta} \,, \\
&&\rm \Pi_\phi =  -\frac{1}{24}\left[\frac{\lambda_1 m_2}{\eta} +24p_\phi\right]+ \frac{\lambda_1 \alpha}{24 \eta}
coth \left(\frac{\alpha}{2}t\right) \,, \\
&&\rm \Pi_\sigma = -\frac{1}{24}\left[\frac{\lambda_2 m_2}{\eta} +24p\phi\right]+ \frac{\lambda_2 \alpha}{24 \eta}
coth \left(\frac{\alpha}{2}t\right) \,,
\end{eqnarray}
where ($\rm \Omega_0, \phi_0, \sigma_0$) are integration constants. In order to fulfill the EKG equations Eqs.(\ref{ein0}-\ref{ein3}), all constants must satisfy that $\rm 1152 \eta V_0=\alpha^{2}e^{\lambda_1 \phi_0 +\lambda_2 \sigma_0-2\eta\Omega_{0}}$. Finally the scale factor $\rm A=e^ \Omega$ becomes
\begin{equation}
\rm A=A_0 e^{\frac{m_2}{2\eta}t} \,
csch^{1/\eta} {\left(\frac{\alpha}{2}t\right)} \,,
\end{equation}
where $\rm A_{0}=e^{\Omega_{0}}$. For this case, given that $\rm\eta>0$, one would expect the scale factor to grow in a similar way as the previous case $\rm\lambda^{2}>3$.

\subsection{Number of e-folds \label{csolutionsd}}

Inflation is characterised by the number of e-folds it expands during such period, that corresponds to $\rm A''_{phys}>0$, where the primes represent the derivatives with respect to the cosmic time $\rm t_{phys}$. The e-folding function $\rm N_{e}=\int dt_{phys}H(t_{phys})$ is described by $\rm t_{phys}$: computing the integral from $\rm t_{phys}*$ to $\rm t_{phys\,\,end}$; where $\rm t_{phys}*$ represents the time when the relevant cosmic microwave background (CMB) modes become superhorizon at 50-60 e-folds before inflation ends at $\rm t_{phys\,\,end}$; and $\rm H(t_{phys})=H_{phys}=A'_{phys}/A_{phys}$ is the Hubble parameter. Although, in our prescription we use a proper time $\rm t$, we can evaluate the Hubble function in the corresponding gauge as $\rm H_{phys}=\dot{\Omega}/N$.

At the end of inflation the expansion rate of the scale factor must be null which translates to $\rm -H'_{phys}=H_{phys}^{2}$ or $\rm\ddot{\Omega}=2\dot{\Omega}^{2}$. From here we can compute the time when inflation ends ($\rm t_{end}$) given each particular case. In Table \ref{t:solutions} appears the computation of the e-folding function $\rm N_{e}$ and $\rm t_{end}$ for each case given by the $\rm\lambda$ parameter.
\begin{center}
\begin{table}[ht]
\begin{scriptsize}
\renewcommand{\arraystretch}{3.5}
\begin{tabular}{|r|c|c|c|}
\hline
 & $\rm t_{end}$ & $\rm M_{\pm}^{(n)}$  &$\rm N_{e}$   \\ \hline
$\lambda=3$ & $\rm\frac{1}{m_{2}}\ln[M^{(1)}_{\pm}]$ & $\rm \frac{1}{8c_{1}}\left[\frac{m_{2}^{2}-8m_{3}}{m_{2}}\pm\sqrt{m_{2}^{2}-16m_{3}}\right]$ & $\rm \frac{2}{m_{2}}\left[\frac{m_{3}}{m_{2}}\ln[M^{(1)}_{\pm}]+c_{1}M^{(1)}_{\pm}-m_{3}t_{*}-c_{1}e^{m_{2}t_{*}}\right]$  \\ \hline
$\lambda>3$ & $\rm\frac{2}{\omega}\tanh^{-1}\left[M^{(2)}_{\pm}\right]$ &$\rm\frac{-2m_{2}\pm\sqrt{\beta[\omega^{2}(2-\beta)+2m_{2}^2]}}{\omega(2-\beta)}$& $\rm\frac{1}{2\beta}\left[\left(\frac{m_{2}+\omega}{\omega}\right)\ln[1+M^{(2)}_{\pm}]-\left(\frac{m_{2}-\omega}{\omega}\right)\ln[1-M^{(2)}_{\pm}]-m_{2}t_{*}-2\ln\left[\cosh\left(\frac{\omega}{2}t_{*}\right)\right]\right]$ \\ \hline
$\lambda<3$ & $\rm\frac{2}{\alpha}\coth^{-1}\left[M^{(3)}_{\pm}\right]$ &$\rm\frac{-2m_{2}\pm\sqrt{\eta[\alpha^{2}(\eta-2)+2m_{2}^2]}}{\alpha(\eta-2)}$& $\rm\frac{1}{2\eta}\left[\left(\frac{m_{2}+\alpha}{\alpha}\right)\ln[M^{(3)}_{\pm}+1]-\left(\frac{m_{2}-\alpha}{\alpha}\right)\ln[M^{(3)}_{\pm}-1]-m_{2}t_{*}+2\ln\left[\sinh\left(\frac{\alpha}{2}t_{*}\right)\right]\right]$\\
\hline
\end{tabular}
\renewcommand{\arraystretch}{1}
\end{scriptsize}
\caption{Computation of the number of e-folds $\rm N_{e}$ and $\rm t_{end}$ for each case provided by the $\rm\lambda$ parameter. Note that $\rm M_{\pm}^{(1)}>1$, $\rm 0<M_{\pm}^{(2)}<1$, and $\rm M_{\pm}^{(3)}>1$ in order to have that $\rm t_{end}>0$.  \label{t:solutions} }
\end{table}
\end{center}

\section{Quantum solutions \label{qsolutions}}

The Wheeler-DeWitt equation for this model is acquired by replacing  $\rm \Pi_{q^\mu}=-i\hbar \partial_{q^\mu}$
in (\ref {hamifrw}).  The factor $\rm e^{-3\Omega}$ may be factor ordered with $\rm \hat \Pi_\Omega$ in several forms. Hartle and Hawking \citep{HH} have suggested what might be called a semi-general factor ordering, which
in this case would order $\rm e^{-3\Omega} \hat \Pi^2_\Omega$ as
\begin{eqnarray}
\rm - e^{-(3- p)\Omega}\, \partial_\Omega e^{-p\Omega} \partial_\Omega&=&\rm - e^{-3\Omega}\, \partial^2_\Omega +
 p\, e^{-3\Omega} \partial_\Omega, \label {hh}
\end{eqnarray}
where $\rm p$ is any real constant that measures the ambiguity in the factor ordering for the variable $\Omega$, in the following we will assume such factor ordering for the Wheeler-DeWitt equation, which becomes
\begin{equation}
\rm \hbar^2 \Box \Psi+ \hbar^2 p\frac{\partial \Psi}{\partial \Omega}- U(\Omega,\phi,\sigma)\Psi=0, \label{wdwmod}
\end{equation}
 where $\rm \Box=-\frac{\partial^2}{\partial
\Omega^2}+\frac{1}{12}\frac{\partial^2}{\partial \phi^2}+\frac{1}{12}\frac{\partial^2}{\partial \sigma^2}$ is the
d'Alambertian in the coordinates $q^\mu=(\Omega,\phi,\sigma)$ and the
potential is $\rm U=  +24V_0 e^{6\Omega-\lambda_1 \phi-\lambda_2 \sigma} $. Then we transform the coordinates to obtain a potential that only depends on a single variable, employing the following transformation
\begin{eqnarray}
\rm \zeta&=&\rm 6\Omega-\lambda_1 \phi-\lambda_2 \sigma,\nonumber\\
\rm \kappa &=&\rm \phi+ \sigma,\nonumber\\
\rm \eta &=& \rm \phi- \sigma.
\end{eqnarray}
Now we find the partial derivatives of $\psi$ with respect to the old coordinates (a, $\phi$, $\sigma$) but in terms of the new variables ($\zeta$, $\kappa$, $\eta$),

\begin{eqnarray}
\rm \frac{\partial \Psi}{\partial \Omega}&=&\rm \frac{\partial \Psi}{\partial \zeta} \frac{\partial \zeta}{\partial \Omega} +
\frac{\partial \Psi}{\partial \kappa} \frac{\partial \kappa}{\partial \Omega}+\frac{\partial \Psi}{\partial \eta} \frac{\partial \eta}{\partial \Omega}=6\*\frac{\partial \Psi}{\partial \zeta}, \nonumber\\
\rm \frac{\partial \Psi}{\partial \phi}&=&\rm \frac{\partial \Psi}{\partial \zeta} \frac{\partial \zeta}{\partial \phi} +
\frac{\partial \Psi}{\partial \kappa} \frac{\partial \kappa}{\partial \phi}+\frac{\partial \Psi}{\partial \eta} \frac{\partial \eta}{\partial \phi}=-\lambda_1 \frac{\partial \Psi}{\partial \zeta}+ \frac{\partial \Psi}{\partial \kappa}+ \frac{\partial \Psi}{\partial \eta}, \nonumber\\
\rm \frac{\partial \Psi}{\partial \sigma}&=& \rm \frac{\partial \Psi}{\partial \zeta} \frac{\partial \zeta}{\partial \sigma} +
\frac{\partial \Psi}{\partial \kappa} \frac{\partial \kappa}{\partial \sigma}+\frac{\partial \Psi}{\partial \eta} \frac{\partial \eta}{\partial \sigma}=-\lambda_2 \frac{\partial \Psi}{\partial \zeta}+\frac{\partial \Psi}{\partial \kappa}-\frac{\partial \Psi}{\partial \eta},
\end{eqnarray}
from here we use these new relations in the quantum Hamiltonian density, obtaining
\begin{eqnarray}
&&\rm 12\hbar^2  \left(\lambda_{1}^2+\lambda_{2}^2-3\right) \frac{\partial^2 \Psi}{\partial \zeta^2}
+24 \hbar^2\left(-\lambda_{1}+\lambda_{2}\right)\frac{\partial^2\Psi}{\partial \eta \partial \zeta}+
24 \hbar^2\frac{\partial^2 \Psi}{\partial \eta^2}-24 \hbar^2 \left(\lambda_{1}+\lambda_{2}\right)  \frac{\partial^2 \Psi}{\partial \kappa \partial \zeta}+ 24 \hbar^2 \frac{\partial^2 \Psi}{\partial \kappa^2}\nonumber\\
&&\rm
-6 \hbar^2 p \frac{\partial \Psi}{\partial \zeta}-24 V_0 e^{\zeta}\Psi=0,
 \label{wdw}
\end{eqnarray}
At this point, we propose the following ansatz, $\rm \Psi= e^{\frac{1}{\hbar}\left( c_2 \kappa
+ c_3\eta \right)}\, G(\zeta),$ where the parameters $\rm c_i$ are constants and
$\rm G(\zeta)$ is a function to be determined. By introducing the aforementioned into Eq.(\ref{wdw}) we obtain the following differential equation of the function $\rm G$,
\begin{equation}
\rm \delta_0 \frac{d^2 G}{d\zeta^2}+ \alpha_0 \frac{dG}{d\zeta}+ \left(\beta_0 e^{\zeta} + \rho_0 \right)G=0, \label{wdw-m}
\end{equation}
where the constants are
\begin{eqnarray}
\rm \delta_0 &=&\rm 12\hbar^2 \left(\lambda_1^2+\lambda_{2}^2-3\right) \,,\nonumber\\
\rm \alpha_0&=& \rm  -6\hbar  \left[ 4\lambda_1(c_2+c_3) + 4\lambda_2(c_2- c_3) + \hbar p\right]\,, \nonumber\\
\rm \beta_0 &=& -24 V_0 \,,\rm  \nonumber\\
\rm \rho_0&=&\rm 24(c_2^2+c_3^2) \,.
\end{eqnarray}

The solution of Eq.(\ref{wdw-m}) is dependant to the value  of constant $\rm \delta_0$, which turns in three different cases, {\bf I)} $\rm \delta=0$ implying that $\rm \lambda_1^2+\lambda_2^2=3$, {\bf II)} $\rm \delta < 0$ implying that $\rm \lambda_1^2+\lambda_2^2 < 3$ and {\bf III)} $\rm \delta > 0$ implying that $\rm \lambda_1^2+\lambda_2^2 > 3$, which can be analyzed in two different cases.

\subsection{case $\delta=0$ \label{qsolutionsa}}
For this case, the Eq.(\ref{wdw-m}) becomes
\begin{equation}
\rm \alpha_0 \frac{dG}{d\zeta}+ \left(\beta_0 e^{\zeta} + \rho_0 \right)G=0 \,, \label{wdw-m2}
\end{equation}
which solution is
\begin{equation}
\rm G(\zeta)=e^{c_1 \zeta}\, Exp[-\frac{4 V_0}{\hbar[4\lambda_1(c_2+c_3)+4\lambda_2(c_2-c_3)+\hbar p]} \, e^{\zeta}],\qquad c_1=\frac{4 (c_2^2+c_3^2)}{\hbar[4\lambda_1(c_2+c_3)+4\lambda_2(c_2-c_3)+\hbar p]}\,,  \end{equation}
therefore, the corresponding wave function for this case becomes
\begin{equation}
\rm \Psi=\psi_0 e^{\frac{1}{\hbar}(c_1\zeta + c_2 \kappa + c_3 \eta)}\, Exp\left[-c_4 \, e^{\zeta} \right] \,, \qquad c_4=\frac{4 V_0}{\hbar[4\lambda_1(c_2+c_3)+4\lambda_2(c_2-c_3)+\hbar p ]} \label{case0} \,.
\end{equation}
Note that wave function has a damping behavior with respect to the scale factor, which is a required feature.

\subsection{case $\delta \not=0$ \label{qsolutionsb}}

For this case, the Eq.(\ref{wdw-m}) becomes, which is similar to that in \cite{polyanin},
\begin{equation}
\rm y^{\prime \prime} + a y^\prime + \left(b e^{\kappa x } +c
\right)y=0 \,, \qquad y=e^{-\frac{ax}{2}} Z_\nu \left(
\frac{2\sqrt{b}}{\kappa} e^{\frac{\kappa x}{2}} \right) \,,
\end{equation}
where $\rm Z_\nu$ is the Bessel function and $\nu=\frac{\sqrt{a^2-4c}}{\kappa}$
the corresponding order, and its relations are
\begin{eqnarray}
\rm a &=&\rm \frac{\alpha_0}{\delta_0}=-\frac{4\lambda_1(c_2+c_3)+4\lambda_2(c_2-c_3)+\hbar p}{\hbar(\lambda_1^2+\lambda_2^2-3)}, \nonumber\\
\rm b &=&\rm \frac{\beta_0}{\alpha_0}=\left\{
\begin{tabular}{ll}
$\rm -\frac{2V_0}{\hbar^2(\lambda_1^2+\lambda_2^2-3)},$ & when \,\,$\lambda_1^2+\lambda_2^2 > 3$\\
$\rm \frac{2V_0}{\hbar^2(\lambda_1^2+\lambda_2^2-3)},$ & when \,\,$\lambda_1^2+\lambda_2^2 < 3$
\end{tabular}
\right.
\nonumber\\
\rm c &=& \rm \frac{\rho_0}{\alpha_0 }=\frac{2(c_2^2+c_3^2)}{\hbar^2(\lambda_1^2+\lambda_2^2-3)}, \qquad  \kappa=1,
\end{eqnarray}
which according to the constant b, the solution to the function $\rm G$ becomes
\begin{eqnarray}
\rm G(\zeta) &=&\rm  e^{\frac{4\lambda_1(c_2+c_3)+4\lambda_2(c_2-c_3)+\hbar p}{2\hbar(\lambda_1^2+\lambda_2^2-3)}\zeta}\,\,\, K_\nu\left(
\frac{2}{\hbar}\sqrt{\frac{2V_0}{ \lambda_1^2+\lambda_2^2-3 }}\,\, e^{\frac{\zeta}{2}} \right), \qquad \lambda_1^2+\lambda_2^2 > 3 \label{k}\\
\rm G(\zeta) &=&\rm  e^{-\frac{4\lambda_1(c_2+c_3)+4\lambda_2(c_2-c_3)+\hbar p}{2\hbar(3-\lambda_1^2+\lambda_2^2)}\zeta}\,\,\, J_\nu\left(\frac{2}{\hbar}\sqrt{\frac{2V_0}{ 3-\lambda_1^2+\lambda_2^2}}\,\,
 e^{\frac{\zeta}{2}} \right), \qquad \lambda_1^2+\lambda_2^2 < 3. \label{j}
\end{eqnarray}
and the resulting wave functions are
\begin{eqnarray}
\rm \Psi &=&\rm  e^{c_5\zeta+c_2 \kappa +c_3 \eta}\,\,\, K_\nu\left(
\frac{2}{\hbar}\sqrt{\frac{2V_0}{ \lambda_1^2+\lambda_2^2-3 }}\,\, e^{\frac{\zeta}{2}} \right), \qquad \lambda_1^2+\lambda_2^2 > 3 \label{k2}\\
\rm \Psi &=&\rm  e^{-c_6\zeta+c_2 \kappa +c_3 \eta}\,\,\, J_\nu\left(\frac{2}{\hbar}\sqrt{\frac{2V_0}{ 3-\lambda_1^2+\lambda_2^2}}\,\,
 e^{\frac{\zeta}{2}} \right), \qquad \lambda_1^2+\lambda_2^2 < 3. \label{j2}
\end{eqnarray}
where the constants are
\begin{equation}
\rm c_5=\frac{4\lambda_1(c_2+c_3)+4\lambda_2(c_2-c_3)+\hbar p}{2\hbar(\lambda_1^2+\lambda_2^2-3)} \,, \qquad
c_6=\frac{4\lambda_1(c_2+c_3)+4\lambda_2(c_2-c_3)+\hbar p}{2\hbar(3-\lambda_1^2+\lambda_2^2)} \,.
\end{equation}

Whilst $\rm c_{5}<0$ and $\rm c_{6}>0$ the wave functions Eqs.(\ref{k2},\ref{j2}) will remain suppressed by the growth of the scale factor. Yielding an expected damped wave function.

\section{Conclusions \label{conclusions}}

We studied a flat Friedmann-Robertson-Walker (FRW) multi-scalar field cosmological model. We introduce the corresponding Einstein-Klein-Gordon (EKG) system of equations and the associated Hamiltonian density. Exact solutions to the EKG system are derived by means of Hamilton's approach where a particular scalar potential of the form $\rm V=V_0 e^{-\lambda_1 \phi - \lambda_2 \sigma }$ was utilized, which gave rise to different cases dependant of the free  parameter $\rm \lambda$, for which the scalar fields, the scale factor and the e-folding function were found. The Hamiltonian density was employed in order to compute the Wheeler-DeWitt (WDW) equation, which was solved by means of a change of variables. An ansatz for the wave function was proposed which in turn allowed us to find the  exact form of the generic function and its constants which was composed by, the aforementioned in terms of the free parameter $\rm \lambda$. We found the model to be rather simple and its solutions to be quite interesting for a model building inflation.

\acknowledgments{ \noindent \noindent This work was partially
supported by CONACYT  167335, 179881 grants. PROMEP grants
UGTO-CA-3. RHJ acknowledges CONACYT for financial support. This work is part of the collaboration within the
Instituto Avanzado de Cosmolog\'{\i}a. Many calculations where done
by Symbolic Program REDUCE 3.8. }

%}%


\begin{thebibliography}{99}
\bibitem[Alan H. Guth, 1981]{guth1981} Alan H. Guth
     {\it Inflationary universe: A possible solution to the horizon and flatness problem,}
     \emph{Phys. Rev. D}
     {\bf 23}, 347 (1981).
\bibitem[A. D. H. Linde, 1982]{linde1982} Andrei D. Linde
     {\it A new inflationary universe scenario: A possible solution of the horizon, flatness, homogeneity, isotropy and primordial monopole problems,}
     \emph{Phys. Lett. B}
     {\bf 108}, 389-193 (1982).
\bibitem[J. D. Barrow \& M. S. Turner, 1981]{turner1981} J. D. Barrow and M. S. 		Turner
     {\it Inflation in the Universe,}
     \emph{Nature}
     {\bf 292}, 35-38 (1981) [doi:10.1038/292035a0].
\bibitem[Alexei A. Starobinsky, 1980]{starobinsky1980} Alexei A. Starobinsky
	{\it A new type of isotropic cosmological models without singularity,}
	\emph{Phys. Lett. B}
	{\bf 91}, 99 (1980).
%\cite{Starobinsky:1979ty}
\bibitem{Starobinsky:1979ty}
A.~A.~Starobinsky,
{\it Spectrum of relict gravitational radiation and the early state of the universe,}
JETP Lett.\  {\bf 30}, 682 (1979)
[Pisma Zh.\ Eksp.\ Teor.\ Fiz.\  {\bf 30}, 719 (1979)].
%%CITATION = JTPLA,30,682;%%
%1313 citations counted in INSPIRE as of 05 Sep 2018
	
%\cite{Mukhanov:1981xt}
\bibitem{Mukhanov:1981xt}
V.~F.~Mukhanov and G.~V.~Chibisov,
{\it Quantum Fluctuations and a Nonsingular Universe,}
JETP Lett.\  {\bf 33}, 532 (1981)
[Pisma Zh.\ Eksp.\ Teor.\ Fiz.\  {\bf 33}, 549 (1981)].
%%CITATION = JTPLA,33,532;%%
%1321 citations counted in INSPIRE as of 05 Sep 2018
	
\bibitem[H. Kodama \& M. Sasaki, 1984]{kodama} H. Kodama and M. Sasaki
	{\it Cosmological Perturbation Theory,}
	\emph{Progress of Theoretical Physics Supplement}
	{\bf 78}, 1-166 (1984) [https://doi.org/10.1143/PTPS.78.1]	
	
\bibitem{bassett} B. A. Bassett, S. Tsujikawa and D. Wands,
    \emph{Rev. Mod. Phys.}
    {\bf{78}}, 537 (2006) [arXiv:0507632].	

%\cite{Planck}
\bibitem{Planck}
P. A. R. Ade et al. (Planck Collaboration), \emph{Planck} 2018 results. X. Constraints on inflation,
[arXiv:1807.06211].
%%CITATION = ARXIV:1807.06211;%%"
%1 citations counted in INSPIRE as of 24 July 2018

\bibitem[John D. Barrow, 1985]{barrow} John D. Barrow
	{\it Slow-roll inflation in scalar-tensor theories,}
	\emph{Phys. Rev. D}
	{\bf 51}, 2729 (1995). 	
\bibitem[A. R. Liddle \& Scherrer, 1998]{andrew1998b}    A.R. Liddle,  and R.J. Scherrer
    {\it Classification of scalar field potential with cosmological scaling solutions,}
  	\emph{Phys. Rev. D}
  	{\bf 59}, 023509 (1998)[https://doi.org/10.1103/PhysRevD.59.023509].
\bibitem[Ferreira \& Joyce, 1998]{ferreira}  P.G. Ferreira \& M. Joyce
    {\it Cosmology with a primordial scaling field,}
    \emph{Phys. Rev. D},
    {\bf 58}, 023503 (1998)[https://doi.org/10.1103/PhysRevD.58.023503].
\bibitem[Copeland et al., 2006]{copeland1} E.J. Copeland, M. Sami and S. Tsujikawa
     {\it Dynamics of dark energy}
     \emph{Int. J. Mod. Phys. D}
     {\bf 15} 1753, (2006) [arXiv:hep-th 0603057].
\bibitem{copeland2} E.J. Copeland, Liddle and D. Wands
     {\it Exponential potentials and cosmological scaling solutions,}
     \emph{Phys. Rev. D}
     {\bf 57} 4686, (1998) [https://doi.org/10.1103/PhysRevD.57.4686].
\bibitem{copeland3} E.J. Copeland, T. Barreiro and N.J. Nunes
     {\it Quintessence arising from exponential potentials,}
     \emph{Phys. Rev. D}
     {\bf 61} 127301, (2000) [https://doi.org/10.1103/PhysRevD.61.127301].



\bibitem[Gianluca Calcagni \& Andrew R. Liddle, 2007]{andrew2007} Gianluca Calcagni and Andrew R. Liddle
     {\it Stability of multifield cosmological solutions,}
     \emph{Phys. Rev. D}
      {\bf 77} 023522,(2008) [https://doi.org/10.1103/PhysRevD.77.023522].
\bibitem[D. S\'{a}ez-G\'{o}mez, 2008]{gomez} D. S\'{a}ez-G\'{o}mez
     {\it Scalar-Tensor theories and current Cosmology,}
     \emph{Problems of Modern Cosmology}
     (2008) [arXiv:0812.1980 (hep-th)].
\bibitem[M. Capone, C. Rubano, P. Scudellaro, 2006]{capone} M. Capone, C. Rubano and P. Scudellaro
     {\it Slow rolling, inflation and quintessence,}
     \emph{Europhys.Lett}
     {\bf 73} 149-155, (2006) [arXiv:astro-ph/0607556].

\bibitem[Kolb \& Turner, 1998]{kolb} E. W. Kolb and M. S. Turner,
    {\it The Early Universe,}
    (Addison-Wesley publishing co., Illinois, 1998). 		
%\bibitem[M.C. Bento et al., 2001]{bento} M.C. Bento, O. Bertolami and N.C. Santos
%     {\it A Two-Field Quintessence Model}
%     \emph{Phys. Rev. D}
%     {\bf 65} 067301, (2001) [arXiv:astro-ph/0106405].

\bibitem[A.A. Coley \& R.J. van den Hoogen, 2000]{coley} A.A. Coley and R.J. van den Hoogen
     {\it The Dynamics of Multi-Scalar Field Cosmological Models and Assisted Inflation}
     \emph{Phys. Rev. D}
     {\bf 62} 023517, (2000) [arXiv:gr-qc/9911075].

%\cite{Copeland:1999cs}
\bibitem{Copeland:1999cs}
  E.~J.~Copeland, A.~Mazumdar and N.~J.~Nunes,
  {\it Generalized assisted inflation},
  {\emph Phys. Rev. D} {\bf 60}, 083506 (1999)
  doi:10.1103/PhysRevD.60.083506
  [astro-ph/9904309].
  %%CITATION = doi:10.1103/PhysRevD.60.083506;%%
  %170 citations counted in INSPIRE as of 14 Sep 2018

%\cite{Yokoyama:2007dw}
\bibitem{Yokoyama:2007dw}
  S.~Yokoyama, T.~Suyama and T.~Tanaka,
  {\it Primordial Non-Gaussianity in Multi-Scalar Inflation},
  {\emph Phys. Rev. D} {\bf 77}, 083511 (2008)
  doi:10.1103/PhysRevD.77.083511
  [arXiv:0711.2920 [astro-ph]].
  %%CITATION = doi:10.1103/PhysRevD.77.083511;%%
  %83 citations counted in INSPIRE as of 14 Sep 2018

  %\cite{Chiba:2008rp}
\bibitem{Chiba:2008rp}
  T.~Chiba and M.~Yamaguchi,
  {\it Extended Slow-Roll Conditions and Primordial Fluctuations: Multiple Scalar Fields and Generalized Gravity},
  {\emph JCAP} {\bf 0901}, 019 (2009)
  doi:10.1088/1475-7516/2009/01/019
  [arXiv:0810.5387 [astro-ph]].
  %%CITATION = doi:10.1088/1475-7516/2009/01/019;%%
  %25 citations counted in INSPIRE as of 14 Sep 2018

%\cite{DeCross}
\bibitem{DeCross}
  M.~P.~DeCross, D.~I.~Kaiser, A.~Prabhu, C.~Prescod-Weinstein and E.~I.~Sfakianakis,
  {\it Preheating after Multifield Inflation with Nonminimal Couplings, I: Covariant Formalism and Attractor Behavior},
  {\emph Phys. Rev. D} {\bf 97}, no. 2, 023526 (2018)
  doi:10.1103/PhysRevD.97.023526
  [arXiv:1510.08553 [astro-ph.CO]];
  %%CITATION = doi:10.1103/PhysRevD.97.023526;%%
  %25 citations counted in INSPIRE as of 18 Sep 2018
 %M.~P.~DeCross, D.~I.~Kaiser, A.~Prabhu, C.~Prescod-Weinstein and E.~I.~Sfakianakis,
  {\it Preheating after multifield inflation with nonminimal couplings, II: Resonance Structure},
  {\emph Phys. Rev. D} {\bf 97}, no. 2, 023527 (2018)
  doi:10.1103/PhysRevD.97.023527
  [arXiv:1610.08868 [astro-ph.CO]];
  %%CITATION = doi:10.1103/PhysRevD.97.023527;%%
  %7 citations counted in INSPIRE as of 18 Sep 2018
 %M.~P.~DeCross, D.~I.~Kaiser, A.~Prabhu, C.~Prescod-Weinstein and E.~I.~Sfakianakis,
  {\it Preheating after multifield inflation with nonminimal couplings, III: Dynamical spacetime results},
  {\emph Phys. Rev. D} {\bf 97}, no. 2, 023528 (2018)
  doi:10.1103/PhysRevD.97.023528
  [arXiv:1610.08916 [astro-ph.CO]].
  %%CITATION = doi:10.1103/PhysRevD.97.023528;%%
  %9 citations counted in INSPIRE as of 18 Sep 2018

%\cite{Hotinli:2017vhx}
\bibitem{Hotinli:2017vhx}
  S.~C.~Hotinli, J.~Frazer, A.~H.~Jaffe, J.~Meyers, L.~C.~Price and E.~R.~M.~Tarrant,
  %``Effect of reheating on predictions following multiple-field inflation,''
  Phys.\ Rev.\ D {\bf 97}, no. 2, 023511 (2018)
  doi:10.1103/PhysRevD.97.023511
  [arXiv:1710.08913 [astro-ph.CO]].
  %%CITATION = doi:10.1103/PhysRevD.97.023511;%%
  %5 citations counted in INSPIRE as of 24 Sep 2018
%%%%%%%%%%%%%%%%%%%%%%%%%%%%%%%%%%%%%%%%%%%%%%%%%%%%%%%%%%%%%%%%	
	
%\bibitem[J.R.L. Santos \& P.H.R.S. Moraes, 2015]{santos} J.R.L. Santos and P.H.R.S. Moraes
%     {\it Fast-roll Solutions from two scalar field inflation}
%     \emph{}
%     (2015) [arXiv:1504.07204 (gr-qc)].

\bibitem[A. R. Liddle et al., 1998]{andrew1998a} Andrew R. Liddle, Anupam Mazumdar and Franz E. Schunck
     {\it Assisted inflation}
     \emph{Phys. Rev. D}
     {\bf 58}, 061301 (1998).

%\bibitem[Juan M.\& J. Socorro, 2013]{juanm} Juan M. Ram\'{i}rez and J. Socorro
%     {\it FRW in Cosmological Self-creation Theory}
%     \emph{Int. J. Theor. Phys.}
%  {\bf 52} 2867-2878, (2013) [arXiv:1206.5413 (gr-qc)].

\bibitem{omar-epjp2017} J. Socorro and Omar E. N\'u\~{n}ez,
    {\it Scalar potentials with multi-scalar fields from quantum cosmology an supersymetric quantum mechanics},
    {\emph Eur. Phys. Journal Plus} {\bf 132}: 168 (2017) [arXiv:1702.00478].

\bibitem[Gibbons \& Gishchuk, 1989]{Gibbons} G.W. Gibbons and L. P. Grishchuk
    {\it Nucl. Phys. B}
    {\bf 313}, 736 (1989).

\bibitem[Zhi, 1987]{Zhi}  Li Zhi Fang and Remo Ruffini, Editors,
    {\it Quantum Cosmology,  Advances Series in Astrophysics and
    Cosmology Vol. 3}
    (World Scientific, Singapore, 1987).

\bibitem[Hartle \& Hawking, 1983]{HH} J. Hartle,  \& S.W. Hawking
    \emph{Phys. Rev. D},
    {\bf 28}, 2960 (1983).

\bibitem[Hawking, 1984]{H} S.W. Hawking
    \emph{ Nucl. Phys. B}
    {\bf 239}, 257 (1984).

\bibitem[Guzm\'an et al., 2007]{wssa}
W.~Guzm\'an, M.~Sabido, J.~Socorro and L.~A.~Ure\~na-L\'opez,
{\it Scalar potentials out of canonical quantum cosmology,}
\emph{Int. J. Mod.  Phys. D } {\bf 16} (4), 641-653 (2007).

\bibitem{nuevo}
J.~Socorro and O.~E.~N\'u\~{n}ez,
{\it Scalar potentials with multi-scalar fields from quantum cosmology an supersymetric quantum mechanics},
{\emph Eur. Phys. Journal Plus} {\bf 132}: 168 (2017) [arXiv:1702.00478]. 	

\bibitem{gssa} W. Guzm\'an, M. Sabido,  J. Socorro and L. Arturo Ure\~na-L\'opez,
    Int. J. Mod.  Phys. D {\bf 16} (4), 641-653 (2007),
    {\it Scalar potentials out of canonical quantum cosmology},
    [gr-qc/0506041]

\bibitem{socorro-doleire} J. Socorro and Marco D'oleire, Phys. Rev. D. {\bf 82}(4) 044008-(1-7) (2010),
    {\it Inflation from supersymmetric  quantum cosmology},  [arXiv:1007.3304].

\bibitem{socorro-pimentel} J. Socorro, Priscila Romero, Luis O. Pimentel and M. Aguero, Int. J. of Theor. Phys. {\bf 52}(8), 2722-2734 (2013), {\it Quintom potentials from quantum cosmology using the FRW cosmological model}  [arxiv:1305.1640].

\bibitem{ssw}
J.~Socorro, M.~Sabido and W.~Ram\'irez and M.~G.~Ag\"uero,
{\it Inflaci\'on cosmol\'ogica vista desde la mec\'anica cu\'antica supersim\'etrica}, Ed. Notabilis Scientia (2013).

\bibitem{polyanin} Zaitsev, V.F., \& Polyanin,  A.D., in {\it Handbook of Exact Solutions for Ordinary Differential Equations}
    (Taylor \& Francis Editorial, 2002).

\end{thebibliography}
\end{document}